# Kinetics and thermodynamics of the degree of order of the *B* cations in double-perovskite $Sr_2FeMoO_6$

T. Shimada, J. Nakamura, T. Motohashi, H. Yamauchi and M. Karppinen*

*Materials and Structures Laboratory, Tokyo Institute of Technology, Yokohama 226-8503, Japan*

* Corresponding author (e-mail: karppinen@msl.titech.ac.jp)

Here we show that the *B*-site cation (Fe/Mo) ordering in the double-perovskite magnetoresistor, $Sr_2FeMoO_6$, is controlled by either kinetics or thermal equilibrium depending on the temperature range. In order to enhance the ordering, long synthesis periods at moderate temperatures are thus required. Employment of a special oxygen-getter-controlled low-$O_2$-pressure encapsulation technique for the sample synthesis enabled us to use long heating periods without evident evaporation of Mo, and thereby reach the thermal equilibrium state and a high degree of *B*-site cation order. For samples fired at 1150 $^{o}$C for long periods ($\geq$ 36 hours), record-high $M_S$ values of 3.7 ~ 3.9 $\mu_B$ were obtained.

## Introduction

Oxide halfmetals (HMs) based on "strongly-correlated" *d* electrons have - since their discovery in the late 1990's - been fascinating condensed-matter scientists, as application of such materials is expected to revolutionize near-future electronics called spintronics.[1] In an ideal HM the carriers are 100 % spin-polarized, which allows us to design tunneling-type magnetoresistance (TMR) devices in which spin-dependent electron transfer across intrinsic or extrinsic barriers is sensitively controlled by external magnetic field. Among the few oxide HMs known to date, ferrimagnetic *B*-site ordered double-perovskite (DP) oxides including the prototype $Sr_2FeMoO_6$ are the most promising magnetoresistors for room-temperature and low-field applications.[2] However, such ordered perovskites characteristically possess a number of material parameters[3,4] including anti-site (AS) defects and anti-phase boundaries that may affect the TMR performance but have not been satisfactorily addressed yet.

The *B*-site ordered DPs, $A_2BB'O_{6-w}$, are derived from the simple perovskite (SP) structure, $ABO_{3-\delta}$, upon co-occupation of the octahedral *B*-cation site with two metal species, *B* and *B*', of different charges, such that in a perfectly ordered DP, each $BO_6$ octahedron is surrounded by six corner-sharing $B'O_6$ octahedra, and *vice versa*. As a rule of thumb, one may assume that the larger the difference between the charges of the two *B*-site cations is, the higher is the equilibrium degree of order at the *B* site.[3,4] For Fe in $Sr_2FeMoO_6$, both $^{57}$Fe Mössbauer spectroscopy[5] and Fe *K*- and *L*-edge XANES data[6] indicate a mixed-valence[7,8] or "valence-fluctuation" state of II/III, having its origin in the fact that the itinerant $4d^1$ electron of formally pentavalent Mo transfers part of its charge and spin density to formally trivalent Fe. In line with the II/III valence state of Fe, an NMR study suggested a mixed-valence state of V/VI for Mo.[9] A saturation magnetization ($M_S$) of 4 $\mu_B$ *per* formula unit is expected on the basis of antiferromagnetic coupling between high-spin $Fe^{2.5}$ ($3d^{5.5}$, $S = 2.25$) and $Mo^{5.5}$ ($4d^{0.5}$, $S = 0.25$), though the magnitude of $M_S$ is not inherent of the mixed-valence concept. [Note that the same value would follow even if one assumes antiferromagnetically coupled $Fe^{III}$ ($3d^5$, $S$ = 2.5) and $Mo^V$ ($4d^1$, $S$ = 0.5), or even $Fe^{II}$ ($3d^6$, $S$ = 2) and $Mo^{VI}$ ($4d^0$, $S$ = 0), the least feasible configuration with ferromagnetic coupling between the $Fe^{II}$ species.] However, the $M_S$ values reported for single-phase $Sr_2FeMoO_6$ samples are typically in the range of 3.0 to 3.7 $\mu_B$,[2,10] being considerably smaller than the predicted value of 4 $\mu_B$. This has been attributed to a certain amount of trivalent Fe atoms being misplaced at the Mo site (and to the same amount of pentavalent Mo at the Fe site), *i.e.* AS defects,[2,9] or at antiphase boundaries.[8] Existence of AS Fe atoms (as well as their trivalent valence state) was clearly visible in the Mössbauer spectra obtained for $Sr_2FeMoO_6$ samples,[5,11] and both Monte Carlo simulations[12] and *ab-initio* band structure calculations[13] concluded that the substantial reduction in $M_S$ is due to AS disorder.

In previous works, strong dependence of the low-field magnetoresistance on the degree of *B*-site cation order in $Sr_2FeMoO_6$ was demonstrated.[14,15] Also shown was that the degree of order crucially depends on the sample synthesis conditions.[9] We have clarified this dependence in detail. In this letter we show that depending on temperature the *B*-site cation order in $Sr_2FeMoO_6$ is controlled either by kinetics or by thermal equilibrium. Furthermore, a recipe to obtain samples with high degrees of order is provided.

## Experimental Section

**Sample Synthesis.** For the synthesis of the $Sr_2FeMoO_6$ samples we employed a special oxygen-getter-controlled low-$O_2$-pressure encapsulation technique,[16] which enabled us to use long heating periods without evident Mo evaporation. Cation-stoichiometric mixture of starting powders of $SrCO_3$, $Fe_2O_3$ and $MoO_3$ was first calcined in air at 900 $^oC$ for 15 hours. The calcined powder was then pelletized, and the pellets were encapsulated together with Fe grains (99.9 % up, under 10 mesh) in a fused-quartz ampoule. The function of the Fe grains is to act as a getter of excess oxygen. Prior to evacuation and sealing, the empty space inside the ampoule was filled with fused-quartz rod(s). For the present study, the synthesis was carried out either for a fixed period of 50 hours at different temperatures ranging from 900 to 1300 $^{o}$C or at 1150 $^{o}$C for various periods ranging from 4 to 150 hours.

**Oxygen Stoichiometry.** Using a few representative samples it was confirmed that the oxygen content of the samples remains constant thorough the synthesis condition range employed. For these oxygen-content analyses, a redox method[16] based on coulometric titration of $Fe^{II}$ and/or $Mo^{V}$ species formed upon acidic dissolution of the sample was applied. For this analysis, the sample was dissolved in oxygen-freed 1 M HCl solution, and the anodic oxidation of $Fe^{II}$ and/or $Mo^{V}$ was performed at a constant current of 3 mA. The end-point titration time was read at 820 mV against an Ag/AgCl electrode.

**Degree of Order.** Phase purity of the synthesized samples was confirmed by x-ray powder diffraction (XRD; MAC Science M18XHF[22]; Cu*Kα* radiation). The XRD profiles were further analyzed using a Rietveld refinement program, RIETAN 2000.[17] All the patterns were readily refined with $R_{wp}$ = 6 ~ 12 % in tetragonal space group, *I*4/*m*.[18,19] In the refinement the individual occupancies of Fe and Mo at the two *B* cation sites were let to freely vary under the constraints that the overall occupancy of the both sites is 100 % and the total amounts of the two metals are in the nominal 1:1 atomic ratio. The degree of *B*-cation order or the long-range order parameter ($S$) was defined by: $S \equiv (\omega_{Fe} - x_{Fe})/(1 - x_{Fe})$, and calculated using the refined occupancy of Fe at the "right" Fe site ($\omega_{Fe}$) and the atomic fraction of Fe among the *B*-site atoms ($x_{Fe}$ = 0.5 in the present case).

**Magnetic Characteristics.** Magnetization measurements were performed for the samples using a dc superconducting quantum interference device (SQUID; Quantum Design: MPMS-XL5). The value of $M_S$ was defined as the observed magnetization (*per* formula unit) at 5 T and 5 K. The ferrimagnetic transition temperature ($T_C$) was determined for some of the samples using a thermobalance (Perkin Elmer: Pyris 1 TGA) equipped with a permanent magnet that was placed below the sample holder.

## Results and Discussion

From x-ray diffraction patterns, no traces of impurity phases were seen for any of the samples synthesized. For the oxygen-getter-controlled low-$O_2$-pressure encapsulation technique used for sample synthesis, the partial pressure of oxygen depends on the synthesis temperature ($T_{syn}$), within $10^{-10} \sim 10^{-16}$ atm for the $T_{syn}$ range of 900 ~ 1300 °C.[20] In order to confirm that the oxygen content of the obtained sample does not vary accordingly we analyzed the precise oxygen of samples synthesized at 900, 1000 and 1150 °C by means of wet-chemical redox analysis. For all the three samples the oxygen content *per* formula unit was determined at 6.00 within ± 0.03. We may thus conclude that our samples used for evaluating the degree of *B*-site cation order are of high quality and internally comparable in terms of both phase purity and oxygen stoichiometry.

In Fig. 1(a), the long-range order parameter, $S$, is plotted against the synthesis temperature, $T_{syn}$, for samples fired for 50 hours at various temperatures ranging from 900 °C to 1300 °C. A maximum in $S$ is seen about 1150 °C, indicating that thermodynamic equilibrium is reached in the vicinity of that temperature. For the temperature range lower than 1150 °C, $S$ increases as $T_{syn}$ increases. This is opposite to the trend expected for systems in which the degree of order is limited by thermodynamics: in this temperature range the ordering is apparently controlled by kinetics. On the other hand, the data in Fig. 1(a) above 1150 °C should correspond to the thermal equilibrium values of $S$ at respective temperatures. In Fig. 1(b), these datum points are fitted to a curve based on the zeroth-order Bragg-Williams approximation ($n$ = 1 in the following formula) and Cowley's theory for the second order phase transition in the β-CuZn-type phase ($n$ = 2):[21] $\ln[(1+S^n)/(1-S^n)] = 2S^n(T_c/T)$, where $T_c$ is a transition temperature (*ca.* 2100 °C for the Bragg-Williams approximation and *ca.* 1800 °C for Cowley's approximation) for the transition from the completely disordered structure ($S$ = 0), *i.e.* SP, to an ordered DP with continuously varying degree of order. Neither of the two (low order) approximations works perfectly in the present case, leaving the order of the present order-disorder phase transition unclear.

In order to affect the thermodynamics of the *B*-site ordering, we consider the following result: from the Rietveld refinements, the unit volume of the sample was found to slightly decrease with increasing *S* (not shown here). This is rather what one expects: in a partially disordered DP each higher-valent cation (here $Mo^{V}$) located at an AS has highly charged cations (here $Mo^{V/VI}$) as the nearest-neighbor cations and therefore it creates extra repulsion in the lattice. This extra repulsion is seen as an expansion of the lattice dimension(s).[4] The fact that the better *B*-site ordering results in the higher-density DP oxide leads us to a possibility that application of an external pressure during sample firing/annealing may push up the thermodynamical limit for the degree of order.

Now we discuss the samples for which the degree of order was kinetically controlled. Besides the samples fired at temperatures lower than 1150 $^{o}$C (for 50 hours), (most of) those fired at 1150 $^{o}$C for different time periods ($t_{syn}$) belong to this category. At 1150 $^{o}$C, the present synthesis procedure facilitates fast formation of the $Sr_2FeMoO_6$ phase with a considerably high degree of order. Even for the sample fired at 1150 $^{o}$C for 4 hours only, all the x-ray diffraction peaks are due to the DP phase and the value of *S* is determined at 0.83. For the sample series synthesized at 1150 $^{o}$C for 4 to 150 hours a general trend of increasing *S* with increasing $t_{syn}$ is evident within the whole $t_{syn}$ range investigated. In Fig. 2, we plot *S* against $t_{syn}^{1/2}$ for these samples: a linear relation is seen after $t_{syn}$ = 16 hours, as expected for a diffusion-controlled process. The highest value of $S = 0.95$ is due to the sample fired at 1150 $^{o}$C for 150 hours. From the curve shown in Fig. 1(a), this *S* value roughly corresponds to the thermal equilibrium state at 1150 $^{o}$C. Thus, in order to enhance the degree of *B*-site cation order in $Sr_2FeMoO_6$, long synthesis periods at moderate temperatures are required. Actually the present synthesis technique is highly beneficial as sample encapsulation prevents possible evaporation of constituent metal(s) during long heat treatments to a minimum extent. In terms of future improvements, atomic-scale mixing of the metal species *prior* to synthesis by means of *e.g.* an appropriate wet-chemical route[22] should be beneficial in facilitating the phase-formation and cation-ordering processes at low temperatures (to achieve the $S$ value as high as possible) within reasonable time periods.

Saturation magnetization, $M_S$, was determined for all the samples using a SQUID magnetometer. The obtained $M_S$ values are given in Fig. 3 against the corresponding $S$ value as calculated from the Rietveld refinement result. A linear dependence between the values of $M_S$ and $S$ is obvious in Fig. 3. The solid line in Fig. 3 is a "theoretical" line for the dependence of $M_S$ on $S$ as obtained from the proposed relation:[10,11] $M_S = 4S$, which is based on a simple model that assumes the ferrimagnetic arrangement among all the neighbors. A reasonable agreement between this simple theory and the experimental data is seen. Here we emphasize that the $M_S$ values higher than 3.7 $\mu_B$ obtained for several samples fired at 1150 $^oC$ for longer periods ($t_{syn} \geq 36$ hours) are (to our best knowledge) the highest values so far reported for the $Sr_2FeMoO_6$ phase. Finally, the ferrimagnetic transition temperature ($T_C$) was determined for some of the samples (varying $T_{syn}$, $t_{syn} = 50$ hours) using a thermobalance equipped with a permanent magnet. The $T_C$ values are plotted in Fig. 4 against $S$. With increasing $S$, $T_C$ increases linearly, obeying the trend predicted from a Monte Carlo simulation.[12]

In summary, it has been shown that in order to control the degree of order of the $B$ cations, *i.e.* Fe and Mo, in $Sr_2FeMoO_6$ factors due to both kinetics and thermal equilibrium need to be taken into account. Based on such considerations, we were able to synthesize highly ordered $Sr_2FeMoO_6$ samples with record-high $M_S$ values of 3.7 ~ 3.9 $\mu_B$.

**Acknowledgments.** This work was supported by Grants-in-aid for Scientific Research (Nos. 15206002 and 15206071) from Japan Society for the Promotion of Science. T. Yamamoto is thanked for his contribution in oxygen-content analysis.

## References

1. Wolf, S. A.; Awschalom, D. D.; Buhrman, R. A.; Daughton, J. M.; von Molnár, S.; Roukes, M. L.; Chtchelkanova, A. Y.; Treger, D. M. *Science* **2001**, *294*, 1488.
2. Kobayashi, K.-I.; Kimura, T.; Sawada, H.; Terakura, K.; Tokura, Y. *Nature (London)* **1998**, 395, 677.
3. Galasso, F. *Structure, Properties and Preparation of Perovskite-Type Compounds* (Pergamon, Oxford, 1969).
4. Woodward, P.; Hoffmann, R.-D.; Sleight, A. W. *J. Mater. Res*. **1994**, *9*, 2118.
5. Lindén, J.; Yamamoto, T.; Karppinen, M.; Yamauchi, H. *Appl. Phys. Lett*. **2000**, *76*, 2925.
6. Karppinen, M.; Yamauchi, H.; Yasukawa, Y.; Lindén, J.; Chan, T. S.; Liu, R. S.; Chen, J. M. *Chem. Mater*. **2003**, in press.
7. Sleight, A. W.; Weiher, J. F. *J. Phys. Chem. Solids* **1972**, *33*, 679.
8. Goodenough, J. B.; Dass, R. I.; *Int. J. Inorg. Mater*. **2000**, *2*, 3.
9. Kapusta, Cz.; Riedi, P. C.; Zajac, D.; Sikora, M.; De Teresa, J. M.; Morellon, L.; Ibarra, M. R. *J. Magn. Magn. Mater*. **2002**, *242-245*, 701.
10. Balcells, Ll.; Navarro, J.; Bibes, M.; Roig, A.; Martínez, B.; Fontcuberta, J.; *Appl. Phys. Lett*. **2001**, *78*, 781.
11. Greneche, J. M.; Venkatesan, M.; Suryanarayanan, R.; Coey, J. M. D. *Phys. Rev. B* **2001**, *63*, 174403.
12. Ogale, A. S.; Ogale, S. B.; Ramesh, R.; Venkatesan, T. *Appl. Phys. Lett*. **1999**, *75*, 537.
13. Saha-Dasgupta, T.; Sarma, D. D. *Phys. Rev. B* **2001**, *64*, 64408.
14. Sarma, D. D.; Sampathkumaran, E. V.; Ray, S.; Nagarajan, R.; Majumdar, S.; Kumar, A.; Nalini, G..; Guru Row, T. N. *Solid State Commun*. **2000**, *114*, 465.
15. García-Hernández, M.; Martínez, J. L.; Martínez-Lope, M. J.; Casais, M. T.; Alonso, J. A. *Phys. Rev. Lett*. **2001**, *86*, 2443.

16. Yamamoto, T.; Liimatainen, J.; Lindén, J.; Karppinen, M.; Yamauchi, H. *J. Mater. Chem*. **2000**, *70*, 2342.

17. Izumi, F.; Ikeda, T. *Mater. Sci. Forum* **2000**, *321*,198.

18. Woodward, P. M. Acta Cryst. B **1997**, *53*, 32.

19. Chmaissem, O.; Kruk, R.; Dabrowski, B.; Brown, D. E.; Xiong, X.; Kolesnik, S.; Jorgensen, J. D.; Kimball, C. W. *Phys. Rev. B* **2000**, *62*, 14197.

20. Richardson, F. D.; Jeffes, J. H. E. *J. Iron Steel Inst.* (London) **1948**, *160*, 261.

21. Warren, B. E. *X-ray Diffraction* (Addison-Wesley, Menlo Park, 1968).

22. Song, W. H.; Dai, J. M.; Ye, S. L.; Wang, K. Y.; Du, J. J.; Sun, Y. P. *J. Appl. Phys*. **2001**, *89*, 7678.

**Figure Captions**

**Fig. 1.** (a) Long-range order parameter, $S$, for samples fired at various temperatures, $T_{syn}$, for 50 hours. The dotted lines that explain the datum points are drawn based on mere eye evaluation. (b) Best fits of the datum points of $T_{syn} > 1150$ °C to $\ln[(1+S^n)/(1-S^n)] = 2S^n (T_c/T)$, with $n = 1$ (Bragg-Williams) and $n = 2$ (Cowley).

**Fig. 2.** Long-range order parameter, $S$, for samples fired at 1150 °C for different time periods, $t_{syn}$.

**Fig. 3.** Saturation magnetization, $M_S$, *versus* long-range order parameter, $S$, for all samples. The broken line is a linear regression of the datum points and the solid line is a theoretical line calculated based on the relation: $M_S = 4S$.[10,11]

**Fig. 4.** Curie temperature, $T_C$, *versus* long-range order parameter, $S$, for samples fired at various temperatures for 50 hours. The broken line is a linear regression of the datum points and the solid line is a theoretical line calculated as: $T_C = (150S + 306)$ °C, based on Monte Carlo simulation.[12]

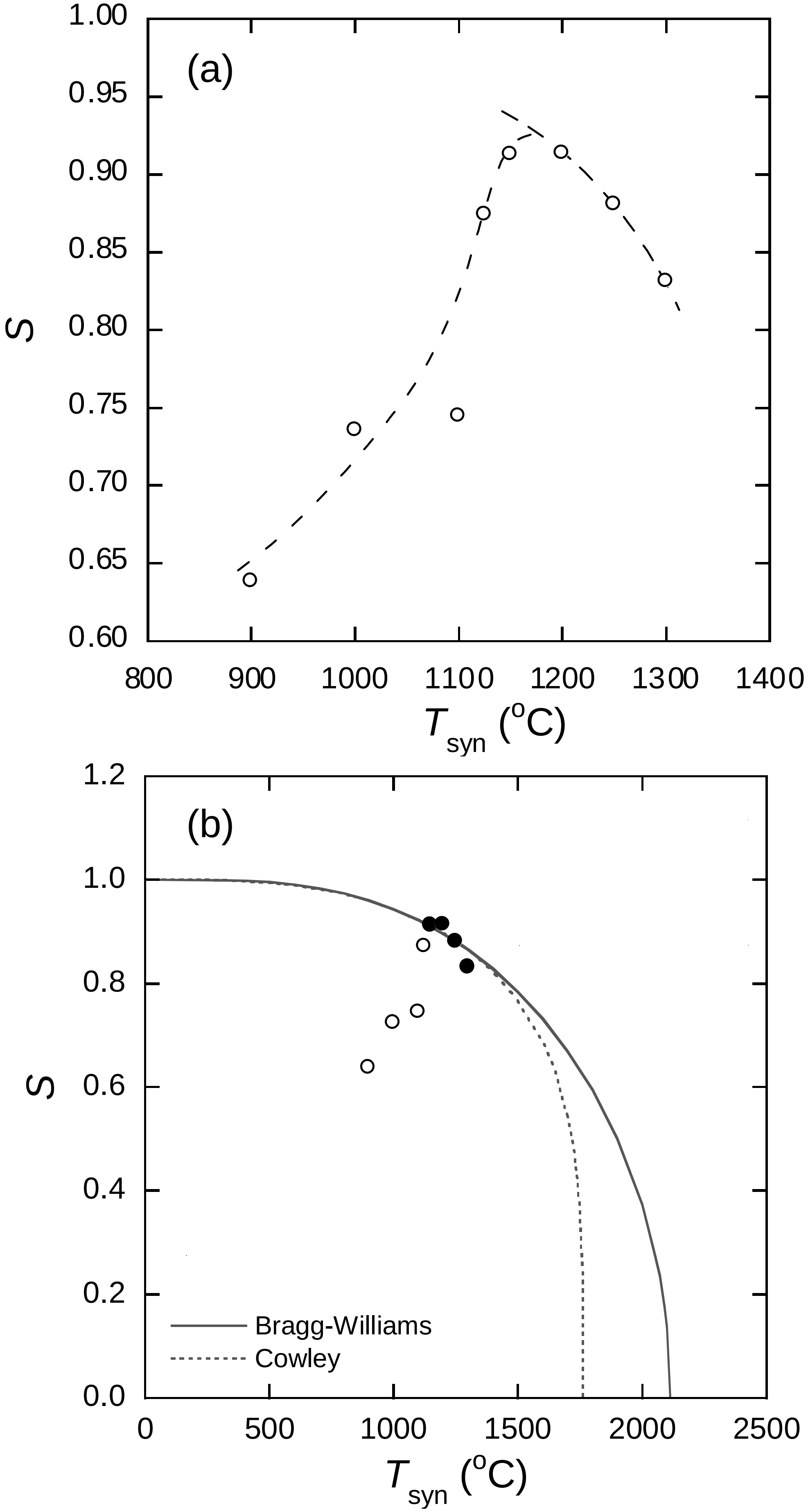

Fig. 1. Shimada *et al*.

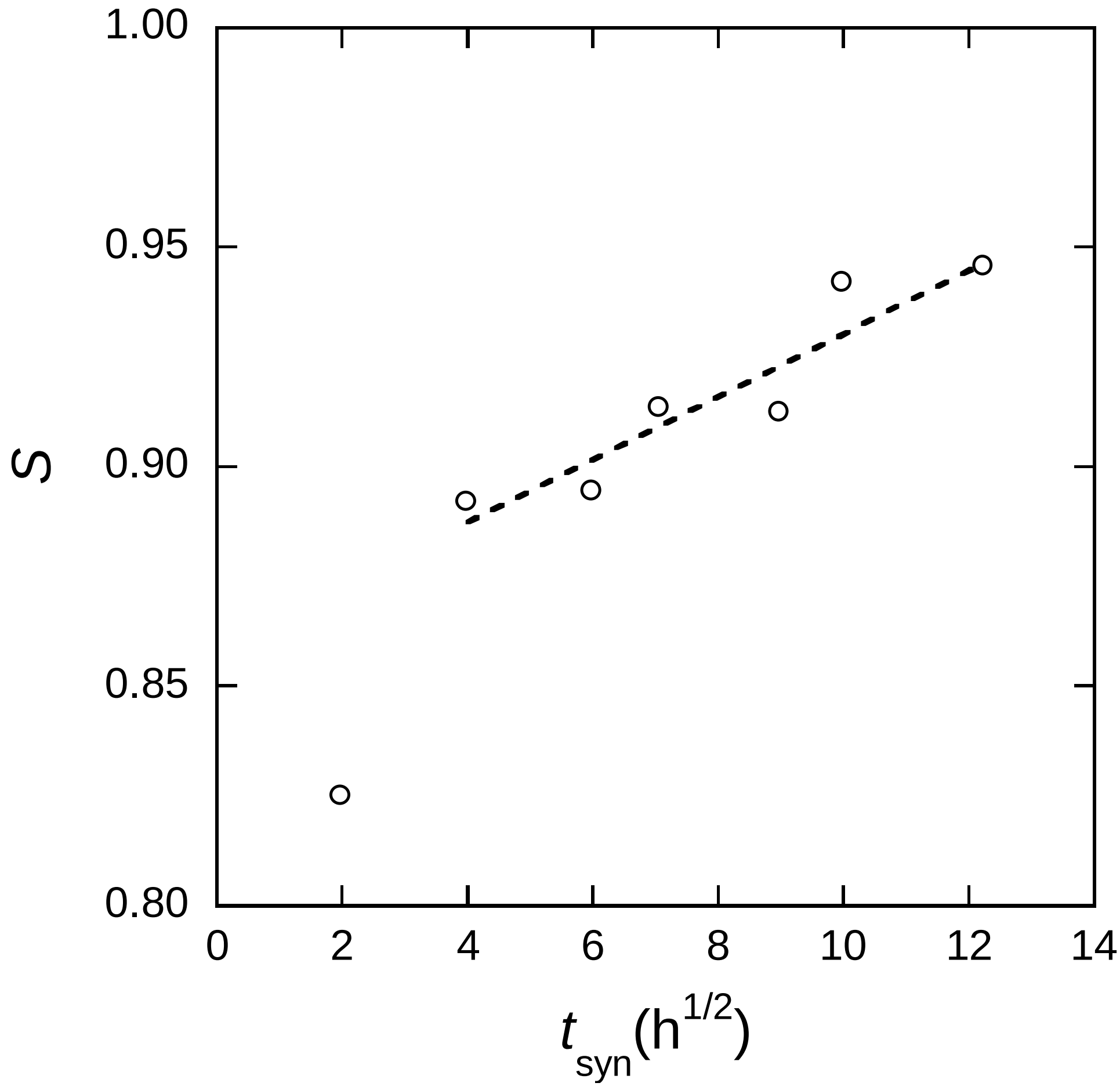

Fig. 2. Shimada *et al*.

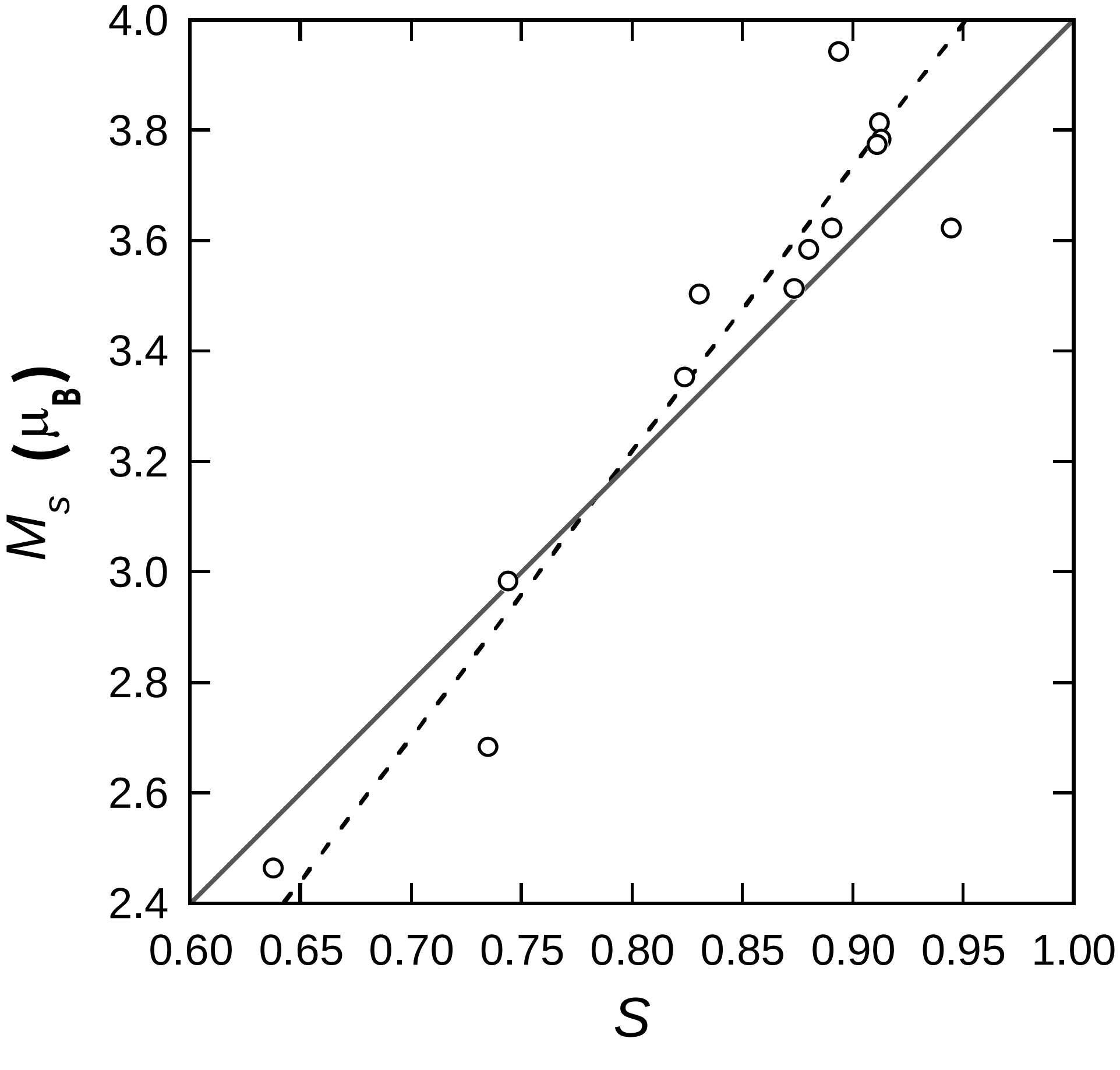

Fig. 3. Shimada *et al*.

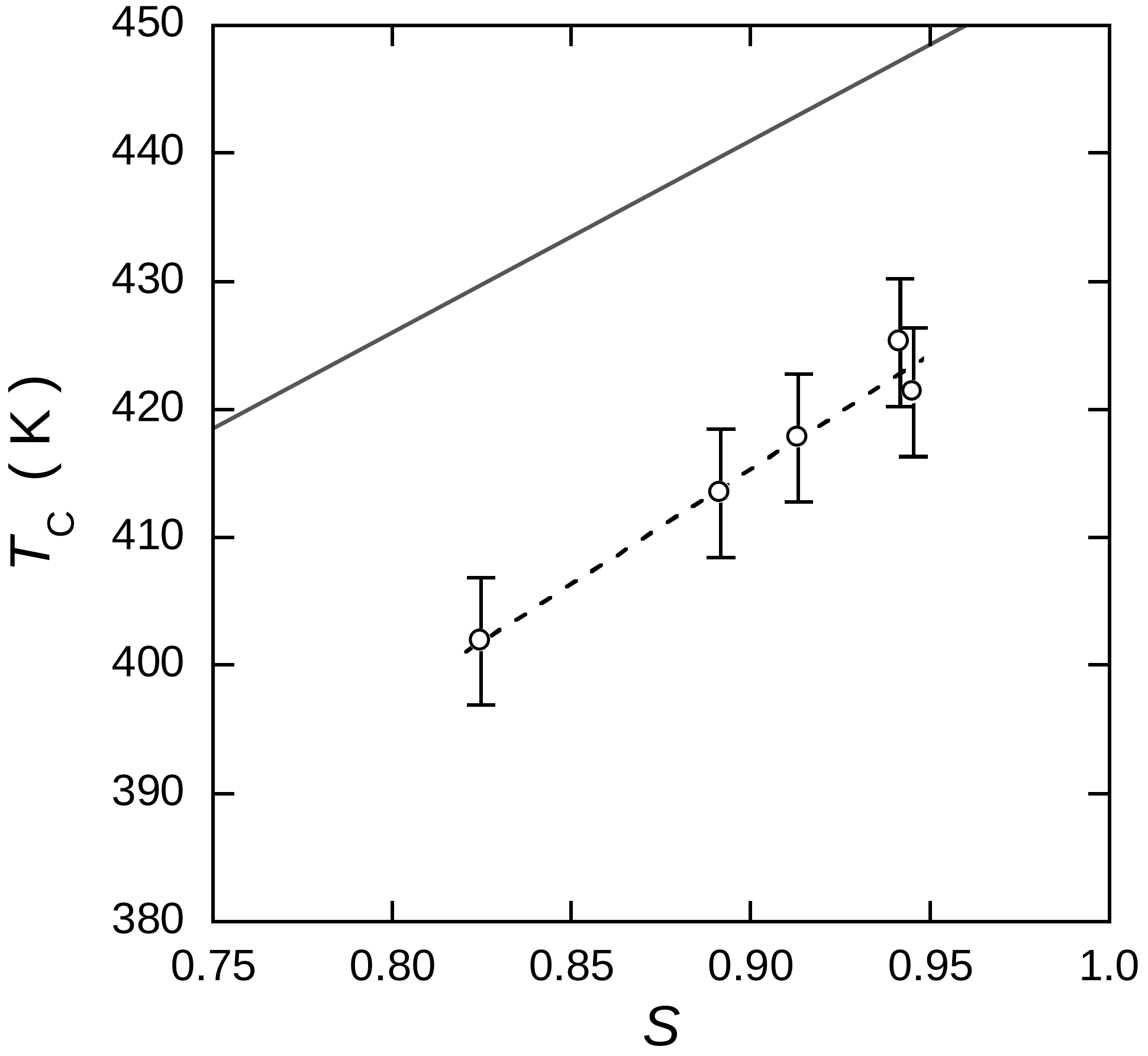

Fig. 4. Shimada *et al*.